\documentclass[12pt]{iopart}
\usepackage{iopams}
\usepackage{graphicx}
\expandafter\let\csname equation*\endcsname\relax
\expandafter\let\csname endequation*\endcsname\relax
\usepackage{amsmath}
\usepackage{amssymb}

\begin{document}

\title[The distribution of path lengths of SAWs on ER networks]
{The distribution of path lengths of self avoiding walks on 
Erd\H{o}s-R\'enyi 
networks}

\author{Ido Tishby, Ofer Biham and Eytan Katzav}
\address{Racah Institute of Physics, 
The Hebrew University, Jerusalem 91904, Israel.}
\eads{\mailto{ido.tishby@mail.huji.ac.il}, \mailto{biham@phys.huji.ac.il}, 
\mailto{eytan.katzav@mail.huji.ac.il}}

\begin{abstract}
We present an analytical and numerical study of   
the paths of self avoiding walks (SAWs) on 
random networks.
Since these walks do not retrace their paths, 
they effectively delete the nodes they visit, 
together with their links,
thus pruning the network.
The walkers 
hop between neighboring nodes,
until they reach a dead-end node from
which they cannot proceed.
Focusing on  
Erd\H{o}s-R\'enyi 
networks we show that the pruned networks maintain a 
Poisson degree distribution,
$p_t(k)$,
with an average degree, 
$\langle k \rangle_t$,
that decreases linearly in time.
We enumerate the SAW paths of any given length and find that
the number of paths, $n_T(\ell)$, increases dramatically as a 
function of $\ell$.
We also obtain analytical results for the path-length distribution, 
$P(\ell)$,
of the SAW paths which are actually pursued, starting from a
random initial node.
It turns out that
$P(\ell)$ follows the Gompertz distribution, which means
that the termination probability of an SAW path
increases with its length.
\end{abstract}


% Random walks, 05.40.Fb
% Graph theory, 02.10.Ox
% Networks in phase transitions, 64.60.aq
% Markov processes, 02.50.Ga
% Random walks and Levy flights, 05.40.Fb
% Fluctuation phenomena, random processes, noise, and Brownian motion, 05.40.-a
% Systems obeying scaling laws, 89.75.Da
\vspace{2pc}

\submitto{\JPA (\today)}
\date
\maketitle

\section{Introduction}
Random walk models 
\cite{Spitzer1964,Weiss1994}
are useful for the study of a large variety of
stochastic processes 
such as Brownian motion and diffusion
\cite{Berg1993,Ibe2013},
polymer structure and dynamics
\cite{Fisher1966,Degennes1979,Doi1986},
and random search
\cite{Evans2011,Lopez2012}.
The models were studied extensively 
in geometries including
continuous space
\cite{Lawler2010b}, 
regular lattices 
of different spatial dimensions
\cite{Lawler2010a},
fractals 
\cite{ben-Avraham2000}
and 
random networks
\cite{Noh2004}.
In the context of complex networks
\cite{Havlin2010,Newman2010}, 
random walks can be used for either probing the
network structure itself
\cite{Costa2007}
or to model dynamical processes such as the
spreading of rumors, opinions and epidemics
\cite{Pastor-Satorras2001,Barrat2012}.
Recent studies focusing on the properties
of random walks on random networks 
have produced analytical results for the
mean first passage time 
\cite{Redner2001}
between two distinct nodes
\cite{Sood2005}, 
the average trapping time, namely the
average time to reach a specific 
node from any starting node 
\cite{Sood2005},  
the average number of distinct nodes 
visited throughout the walk
\cite{Debacco2015}
and the average cover time
\cite{Kahn1989}.

A special type of random walk which has been studied 
extensively on regular lattices is the self avoiding walk
(SAW),
also referred to as the kinetic growth self-avoiding walk
\cite{Herrero2007},
or true or myopic self-avoiding walk
\cite{Slade2011}.
This is a random walk which does not 
visit the same node more than once
\cite{Madras1996}.
At each step, the walker chooses its next move
randomly from the neighbors 
of its present node, excluding nodes
that were already visited.
The walk terminates when it
reaches a dead end node, 
namely a node which does not have any yet unvisited neighbors.
The length of the path, 
$\ell$,
is given by the number 
of steps made until the walk is terminated.
A large number of studies of SAWs on regular lattices were devoted 
to the enumeration of paths as a function 
of their length
\cite{Fisher1959,Kesten1963,Kesten1964,Hara1993,Clisby2007,Clisby2013}.
These studies provided much
insight on the structure and thermodynamics of polymers
\cite{Degennes1979,Doi1986}.
However, SAWs on networks have not attracted much attention
\cite{Herrero2003,Herrero2005,Herrero2005b,Herrero2007,Viana2012}.

SAWs on networks may describe agents or robots 
propagating and damaging a network of computers, 
such that damaged nodes are effectively wiped out from the network.
The path length of an SAW on a connected network of size $N$
can take values between $1$ and $N-1$.
The latter case corresponds to 
a Hamiltonian path
\cite{Bollobas2001}.
More specifically, the SAW path lengths between 
a given pair of nodes,
$i$ and $j$, are distributed in the range bounded from
below by the shortest path length between these nodes
\cite{Katzav2015}
and 
from above by
the longest non-overlapping path between them
\cite{Karger1997}.
From a theoretical point of view, the SAW path length corresponds
to the attrition length, 
also referred to as the
{\it last hitting time}
of the SAW
\cite{Herrero2005}.
This is in contrast to the
{\it first hitting time}
of random walks on networks
\cite{Debacco2015}.
Both studies show that 
the behavior of random walks on random networks exhibits 
common properties with those of random walks on regular lattices of
high dimension.
In particular, they are highly effective in exploring the
space without retracing their steps, in contrast to the
case of low dimensional lattices
\cite{Montroll1965}.

In this paper we study SAWs on 
Erd\H{o}s-R\'enyi (ER)
networks
\cite{Erdos1959,Erdos1960,Erdos1961},
above the percolation threshold.
These walks can be viewed as
random walks which delete the nodes
they visit, thus reducing the network size
one node at a time.
Surprisingly, we find that the remaining network is still an ER network
with connectivity that decreases linearly in time.
We enumerate all the possible SAW paths, 
$n_T(\ell)$, 
of any given length, $\ell$
on an ER network,
providing closed form expressions.
We also study the distribution
$P(\ell)$
of the path lengths of the SAWs
which are actually pursued,
starting from a random node
in the network. 
The analytical results are found to be in excellent agreement 
with numerical simulations.

The paper is organized as follows.
In Sec. 2 we present the SAW model on an ER network.
In Sec. 3 we study the evolution of the network structure as
it is pruned by the SAW.
In Sec. 4 we enumerate the SAW paths.
In Sec. 5 we study the distribution of path lengths.
In Sections 6,7 and 8 we present central measures, 
dispersion measures and extreme value statistics 
of this distribution.
The results are summarized and discussed in Sec. 9.

\section{The self avoiding walk}

Consider a random walk on a random network
of $N$ nodes.
Each time step the walker chooses randomly one of the neighbors
of its current node, and hops to the chosen node. 
Such walks can go on without limit, visiting many nodes multiple times.
Many interesting questions have been studied in this context.
For example, the number of distinct nodes visited by the random walker
as a function of the path length has been calculated in Ref.
\cite{Debacco2015}.
A related property is the average cover time,
which is the average number of steps required for the random
walk to visit all the nodes in the network at least once
\cite{Kahn1989}.
Another interesting question is how many steps the walker makes
until the first time it enters a node which was already visited.
This time is referred to as the first hitting time,
while the path length up to this point is referred to as the
first intersection length.
The distribution of first hitting times was recently studied 
using the cavity approach
\cite{Debacco2015}.

Self avoiding walkers 
are random walkers which
hop only to neighboring nodes which 
have not been visited before. 
%As the walk progresses, the number
%of unvisited neighbors is gradually reduced, limiting the possible moves.
Here we study SAWs on random networks. 
Since the nodes already visited become inaccessible,
the self avoidance condition is equivalent to a process in
which the walkers delete the nodes they visit.
More precisely, the visited node is deleted promptly after the walker
has moved to the next node. 
Thus, the network size is reduced by $1$
at each step.
The edges connected to the deleted
node are also removed.
As a result, the degree of each node which was connected to the
visited node 
is reduced by $1$.
Eventually, the walker may reach a node which does not have any unvisited
neighbors. 
At this point, the path of the random walker is terminated.
We choose as the initial network an $ER(N,p)$ network,
namely
an ER network which consists of $N$ nodes, where
each pair of nodes is connected with probability $p$.
The SAW path starts from a randomly chosen node, 
which is not isolated.
The path length, $\ell$, is given by the number of steps taken
until it terminates.

\section{Evolution of the network structure}

Consider an $ER(N,p)$ network.
The degree $k_i$ of node $i=1,\dots,N$ 
is the number of links connected to this node.
The degree distribution 
$p(k)$
of the ER network 
is a binomial distribution, 
which in the sparse limit
($p \ll 1$)
is approximated by a Poisson distribution of the form 

\begin{equation}
p(k)=\frac{{c}^{k}}{k!}e^{-c},
\label{eq:poisson}
\end{equation}

\noindent
where
$c=(N-1)p$
is the average degree.
In the asymptotic limit 
($N \rightarrow \infty$), 
the ER network exhibits a phase transition 
at $c=1$ (a percolation transition), such
that for $c<1$ the network consists only of small 
clusters and isolated nodes, while for
$c>1$ there is a giant cluster which includes a macroscopic fraction of 
the network, in addition to the small clusters and isolated nodes. 
At a higher value of the connectivity, namely 
at $c = \ln N$, 
there is a second transition, above which the entire
network is included in the giant cluster and there are no isolated
components.
Here we focus on the regime above the percolation transition,
namely $c>1$, where the network includes a giant component.
 
Considering the SAW as a node deleting walk, 
it effectively tears up the network, 
removing one node and its associated links at
every step. 
Clearly, the network size after $t$ steps is
given by
$N(t)=N-t$. 
The degree distribution evolves in time and is 
denoted by
$p_t(k)$, $k=0,\dots,N(t)-1$,
where
$p_0(k)=p(k)$.
The average degree 

\begin{equation}
\langle k \rangle_t = \sum_{k=0}^{N(t)-1} k p_t(k)
\label{eq:<k>}
\end{equation}

\noindent
evolves accordingly.
We denote it by
$c(t) = \langle k \rangle_t$,
where $c(0)=c$.

For random walks on random networks,
there is a higher probability for the walker to
visit nodes with high degrees. 
More precisely, 
the probability to visit a node 
of degree $k$ in the next step is given by  
$k p_t(k)/c(t)$,
namely it is proportional to the 
degree of the node.
A special property of the 
Poisson distribution is that
the probability
${k p_{t}(k)}/{c(t)} = p_{t}(k-1)$.
This means that,
in fact, the probability of a 
node of degree $k$ to be visited
by the random walker 
is proportional to 
$p_{t}(k-1)$. 
%However, by the time the
%node is visited, the previous node 
%is removed, together with the link 
%along which the walker entered the new node.
However, by the time the walker enters the next node,
the previous node is deleted,
together with the edge connecting the two nodes.
Therefore, when the walker enters a node of degree $k$, 
the degree of this node is reduced to $k-1$.
The outcome of this reasoning is that
the probability of 
the SAW to visit
a node of degree $k$ at time $t$ is simply $p_t(k)$, as if
it makes a random choice
of a node in the smaller network.

We now examine the evolution of the network in terms
of the average number of links that 
are removed at each step. 
Deleting a node along the SAW path removes, on average,
$c(t)$ edges, namely $2 c(t)$ half-edges from the node and 
its neighbors. 
The average degree of the network at time $t$ is given by

\begin{equation}
c(t) = \frac{\sum \limits_{i=1}^{N} k_{i}(t)}{N(t)},
\label{eq:c(t)}
\end{equation}

\noindent
where 
$k_i(t)$
is the degree of node $i$ at time $t$,
and $k_i(0)=k_i$.
The degrees of all the nodes already deleted are counted
as $k_i(t)=0$.
The average degree can be expressed as

\begin{equation} 
c(t) = \frac{ \sum \limits_{i=1}^{N} k_{i} 
\left(0\right)-2 \sum \limits_{t^{\prime}=0}^{t-1}
c\left(t^{\prime}\right)}
{N-t}.
\end{equation}

\noindent
Note that
$\sum_{i} k_{i}(0)=N c$.
Therefore, we obtain the recursion equation

\begin{equation}
c(t) = \left(1-\frac{1}{N-t}\right) c(t-1).
\end{equation}

\noindent
Solving this equation we obtain

\begin{equation}
c(t) = \left(1-\frac{t}{N-1}\right)c.
\label{eq:coft}
\end{equation}

\noindent
The correctness of  
Eq.
(\ref{eq:coft})
can also be demonstrated by considering
the case of the complete network, 
$ER(N,p=1)$.
In this case, the SAW visits the entire network with probability
$1$, and 
$c(t) = N(t)-1$ 
for all values of $t$.
This result is consistent with 
Eq.
(\ref{eq:coft}),
where for a complete network 
$c=N-1$.

\begin{figure}
\centerline{
\includegraphics[width=8cm]{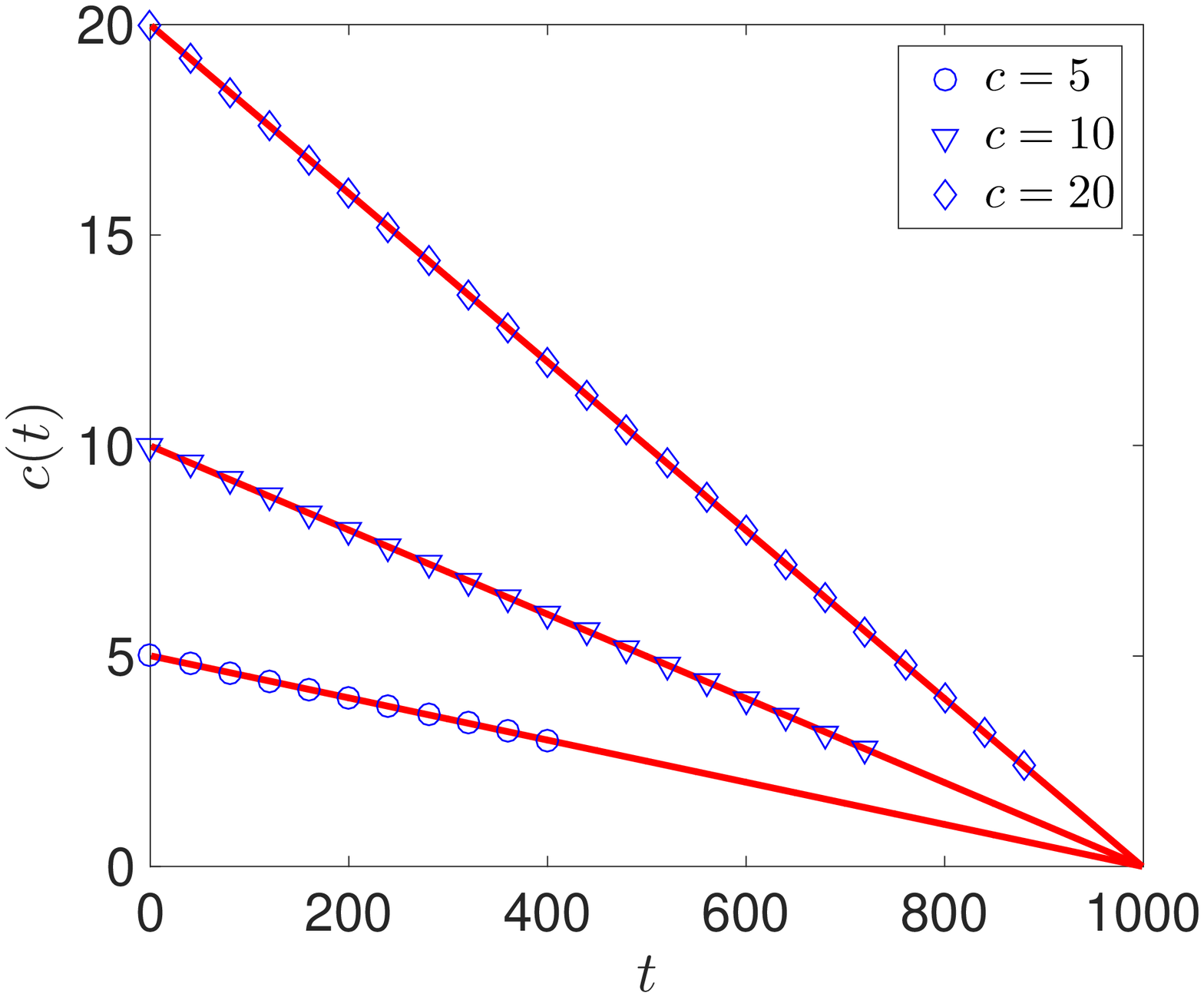}
}
\caption{
The average degree, $c(t)$, of the remaining network vs. time, $t$, 
for an ER network of initial size $N=1000$ and for different initial 
values of the average connectivity, $c$. 
The analytical results (solid lines), obtained from 
Eq. 
(\ref{eq:coft}), 
are in excellent agreement with numerical simulations
for $c=5$ (circles),
for $c=10$ (triangles)
and
for $c=20$ (diamonds).
The numerical results were obtained from $10^4$ realizations
of the random walk for each value of $c$.
} 
\label{fig:c(t)}
\end{figure}

\begin{figure}
\centerline{
\includegraphics[width=8cm]{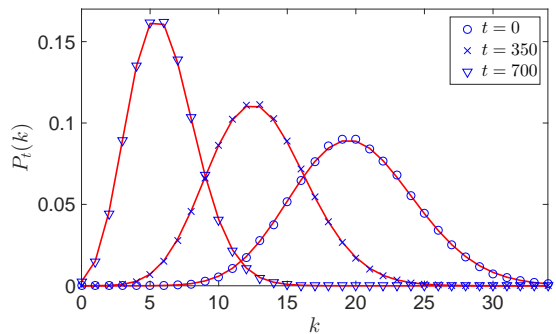}
}
\caption{
The evolution of the degree distribution $p_t(k)$ at 
three different
times, 
$t=0$, $350$ and $700$ 
(represented by circles, crosses and triangles respectively),
on a network of size $N=1000$. 
The initial value of the average degree is $c=20$.
The analytical results (lines), obtained from a Poisson distribution 
[Eq.
(\ref{eq:p_t(k)})],
using the predicted value of $c(t)$ presented in Eq.
(\ref{eq:coft}),
are found to be in excellent agreement with numerical 
simulations.
} 
\label{fig:p_t(k)}
\end{figure}

In Fig. 
\ref{fig:c(t)} 
we present the average degree $c(t)$ vs. time 
for $N=1000$ and different initial values of $c$.
The results obtained from
Eq.
(\ref{eq:coft})
are compared to numerical simulations,
finding excellent agreement.
We now extend the discussion to the temporal evolution of the
entire degree distribution
$p_t(k)$.
In Fig. 
\ref{fig:p_t(k)} 
we present the degree distribution 
$p_t(k)$ vs. $k$,
for $t=0, 350$ and $700$, 
where the initial network is 
$ER(1000,20/1000)$.
Clearly, the degree distribution of the initial network is a 
Poisson distribution with $c=20$
[Eq. (\ref{eq:poisson})].
Interestingly, the degree distribution 
$p_t(k)$
remains a Poisson distribution 
and its average degree
$\langle k \rangle_t$
coincides with $c(t)$ given by Eq.
(\ref{eq:coft}).

In order to understand these results we digress to the simpler 
process of random node deletion, 
in which at each time step a randomly chosen
node is deleted. 
Thus, the probability that the 
deleted node at time $t$ has degree $k$ is given
by $p_t(k)$.
The random node deletion actually maintains the ER character of the 
network, with a Poisson degree distribution and the same value of $p$. 
This property can be easily understood from the fact that
the ER network
can be constructed by starting from a single node
and at each time step adding one node and connecting it to any
existing node with probability $p$
\cite{Bollobas2001}.
Repeating this node addition step $N-1$ times we obtain an ER network
of $N$ nodes.
The node deletion process is simply the time reversal of this construction.

In random node deletion, 
the probability $p$ remains unchanged and thus
$c(t)=[N(t)-1]p$.
Since 
$p=c/(N-1)$ 
and
$N(t)=N-t$,
we find that in random node deletion

\begin{equation}
c(t) = \frac{N(t)-1}{N-1} c, 
\label{eq:RNc(t)}
\end{equation}

\noindent
which coincides with Eq.
(\ref{eq:coft}), describing
SAW deletion.
However, these two processes are different.
The fact that the expressions for $c(t)$ coincide is
a result of two opposing effects which cancel each other.
On the one hand, the probability of a node of degree $k$
to be visited by the
SAW is proportional to its degree
and given by
$k p_t(k)/c(t)$.
Therefore, highly connected nodes are more likely to be removed
from the network compared to the random deletion process.
On the other hand, once the SAW enters a node the link along
which it entered is already deleted, reducing the degree of the
node by $1$.
As mentioned above, it so happens that in the case of a Poisson distribution
$k p_t(k)/c(t) = p_t(k-1)$,
so the net result is that a node with degree $k$ is visited
with probability $p_t(k)$.
The conclusion is that although the SAW and the random deletion are
two different processes, the degree distribution of the remaining 
network is the same.
Therefore, at all times, the degree distribution is 

\begin{equation}
p_t(k) = \frac{c(t)^k}{k!}e^{-c(t)},
\label{eq:p_t(k)}
\end{equation}

\noindent
where $c(t)$ is given by
Eq.
(\ref{eq:coft}).

When node removal processes 
such as random deletion or node-deleting walks
are inflicted on a network with $c>1$,
they drive the network towards the percolation transition.
This transition takes place
at time $t_p$ for which
$c(t_p)=1$. 
using Eq. 
(\ref{eq:coft}) 
one can evaluate the 
time $t_p$,
which is given by
$t_p = (N-1)(1-1/c)$.
It is important to note that in the 
process of random node deletion the percolation
transition is always reached. On the other hand, the SAW path is likely to
terminate long before the percolation threshold is reached. In fact, as the
network approaches the percolation transition, the termination rate of the
SAW paths quickly increases.

Random node deletion and node-deleting walks can be considered
in the context of network attacks. While random node deletion is
an example of a random attack which has been studied extensively
in the literature
\cite{Havlin2010}, 
the node-deleting walk belongs to the
class of localized attacks.
A model of localized attacks which has been studied
in Ref.
\cite{Shao2015},
the attack is initiated at a random node and deletes entire shells of 
neighbors around the initial node, one shell at a time.
It was found that properties of percolation in this model on the ER
network are identical to those obtained by random removal of nodes.
It turns out that node deleting walks on ER networks share this property.
In other networks, such as the scale-free network or the
random regular graph, localized
attacks affect the network differently than random attacks
\cite{Shao2015}.

\section{The number of SAW paths}

In this section we study the combinatorics of the SAW paths
starting from a random node, $i$.
More precisely, we are interested in the  
expected
number of possible
SAW paths of length $\ell$ 
(not necessarily terminating),
which will be denoted by 
$n(\ell)$,
where 
$\ell=1,\dots,N-1$.
Starting at node $i$, 
in the $ER[N,c/(N-1)]$ network,
the initial step of the SAW can be chosen from the
$k_i$ nearest neighbors of $i$.
The average number of nearest neighbors is
$\langle k \rangle = c$.
Hence, $n(1)=c$.
As the SAW proceeds, the network remains an 
ER network with a decreasing value of 
the mean degree
$c(t)$,
given by Eq.
(\ref{eq:coft}).
Therefore, each step, $t$, can be considered as the first SAW step
on the smaller network, 
$ER[N(t),c(t)/(N(t)-1)]$.
As a result, the number of SAW paths of length $\ell$
is simply

\begin{equation}
n(\ell) = \prod_{t=0}^{\ell-1} c(t),
\end{equation}

\noindent
where $c(0)=c$.
Plugging in the expression for $c(t)$ from Eq.
(\ref{eq:coft})
we obtain

\begin{equation}
n(\ell) = 
\frac{(N-1)!}{(N-1-\ell)!} \left( \frac{c}{N-1} \right)^\ell.
\label{eq:n(ell)}
\end{equation}

\noindent
Identifying 
$p=c/(N-1)$ 
and rearranging the last expression yields

\begin{equation}
n(\ell) = 
\binom{N-1}{\ell} \ell! p^\ell.
\label{eq:n(ell-r)}
\end{equation}

\noindent
Expressing $n(\ell)$ in this form highlights the fact that
it accounts for 
all the possible ordered choices of $\ell$ nodes from the 
$N-1$ nodes in the network, 
apart from the initial node, $i$.
The combinatorial factor is
multiplied by the 
probability that all the links along such a 
path exist, which is given by $p^\ell$. 
Using the Stirling approximation we 
obtain 

\begin{equation}
n(\ell) = 
\left( \frac{N}{N-\ell} \right)^{N-\ell+1/2} 
\left( \frac{c}{e} \right)^\ell.
\label{eq:n(ell2)}
\end{equation}

\noindent
For short SAW paths, for which 
$\ell \ll N$, 
this can be further approximated by

\begin{equation}
n(\ell) = 
 e^{-\frac{\ell^2}{2N} + \ell \ln c }.
\label{eq:n(ell3)}
\end{equation}

\noindent
In the limit $N \rightarrow \infty$
this expression reduces to

\begin{equation}
n(\ell) = c^{\ell}.
\label{eq:nred(ell3)}
\end{equation}

\noindent
This resembles the results obtained 
for SAWs on infinite regular lattices
of a finite dimension, $D$,
in which the number of SAW paths on length $\ell$ is

\begin{equation}
n(\ell) \sim \mu^{\ell} \ell^{\alpha},
\label{eq:lattice}
\end{equation}

\noindent
where
$\mu$ is the connective constant
and the exponent $\alpha$
provides a sub-leading correction 
\cite{Fisher1959}.
It was found that 
for $D < 4$
the exponent
$\alpha>0$,
while for 
$D \ge 4$
it satisfies
$\alpha=0$.
Comparing Eqs.
(\ref{eq:nred(ell3)})
and
(\ref{eq:lattice})
we conclude that an SAW on an ER network
is consistent with $\alpha=0$,
and thus resembles an SAW on a regular lattice of 
dimension $D \ge 4$.
On regular lattices, the connective constant
satisfies
$\mu \le z-1$,
where $z$ is the coordination number of the lattice
(for the hyper-cubic lattice $z=2D$).
This is due to the fact that the SAW does not backtrack its path.
In high dimensions the connective 
constant 
$\mu \rightarrow z-1$.
A similar result is obtained for a regular graph in which all
nodes are of degree $c$.
Interestingly, in the ER network the number of paths of length
$\ell$ scales like $c^{\ell}$, where $c$ is the average coordination
number. This is different from the case of regular lattices 
of high dimension,
where
$n(\ell)$ scales like $(z-1)^{\ell}$.
The reason is that the SAW path tends to visit nodes of high degree
more often, in a way that compensates for the loss of the 
backtracking link.

\begin{figure}
\centerline{
\includegraphics[width=5cm]{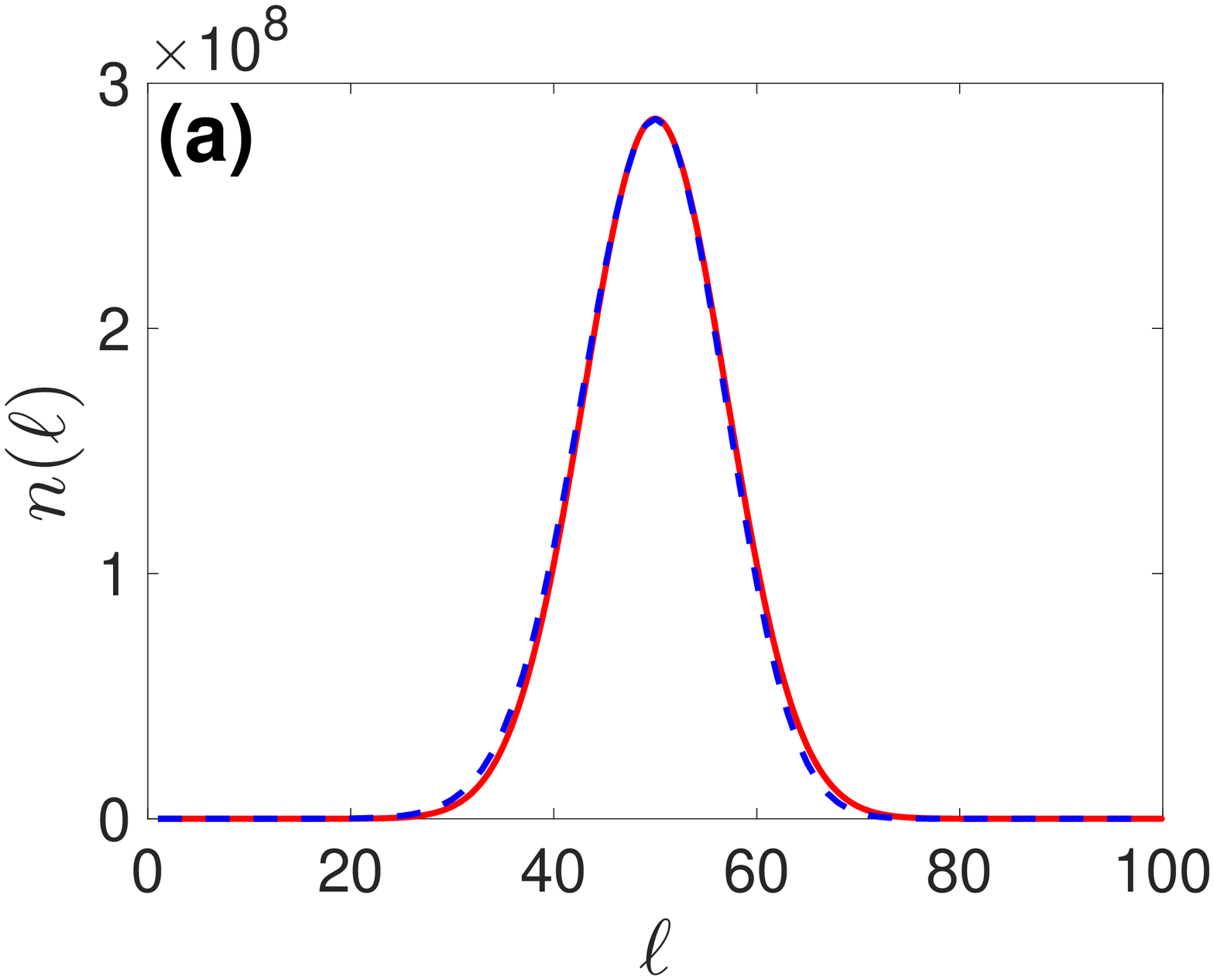}
\includegraphics[width=5cm]{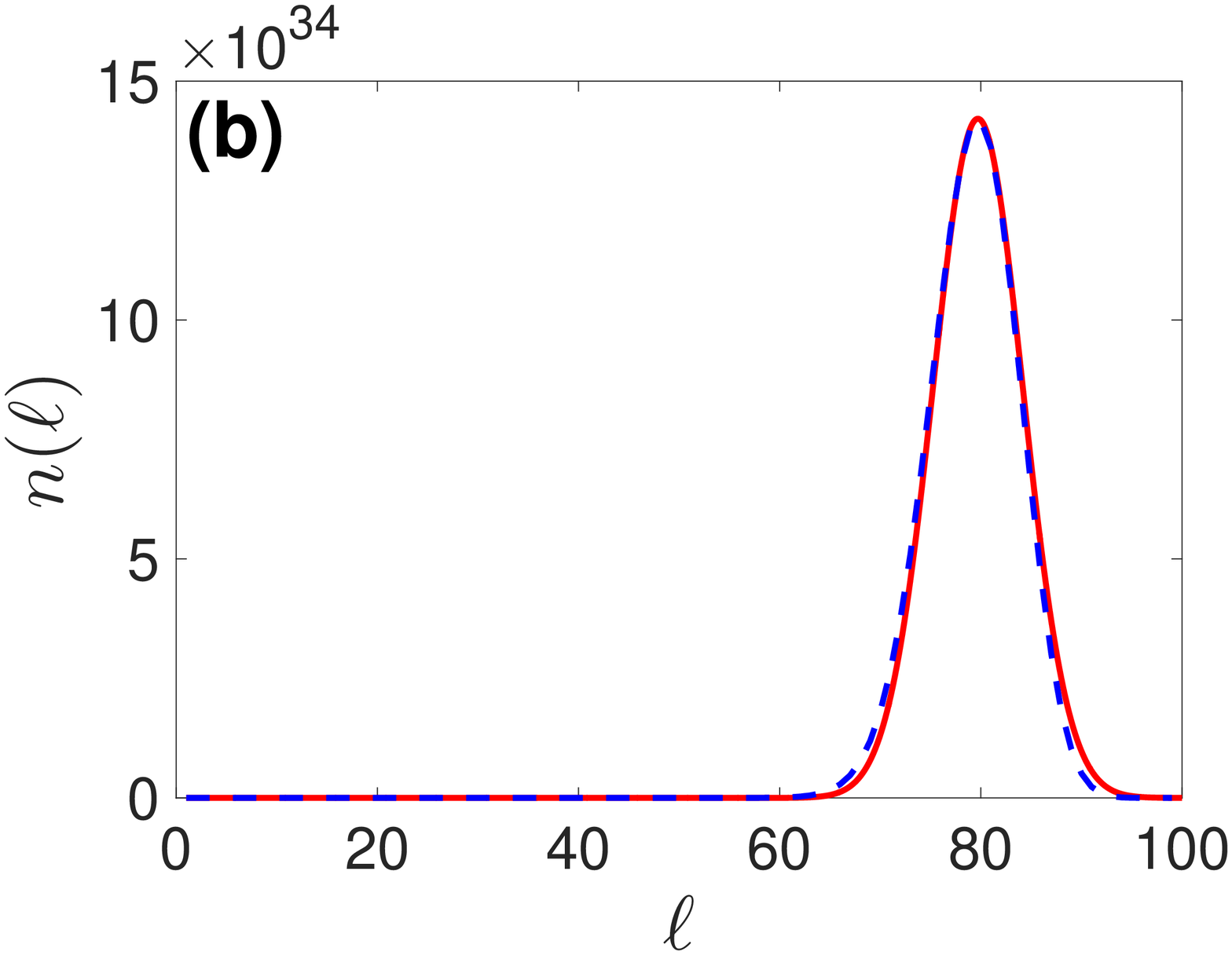}
\includegraphics[width=5cm]{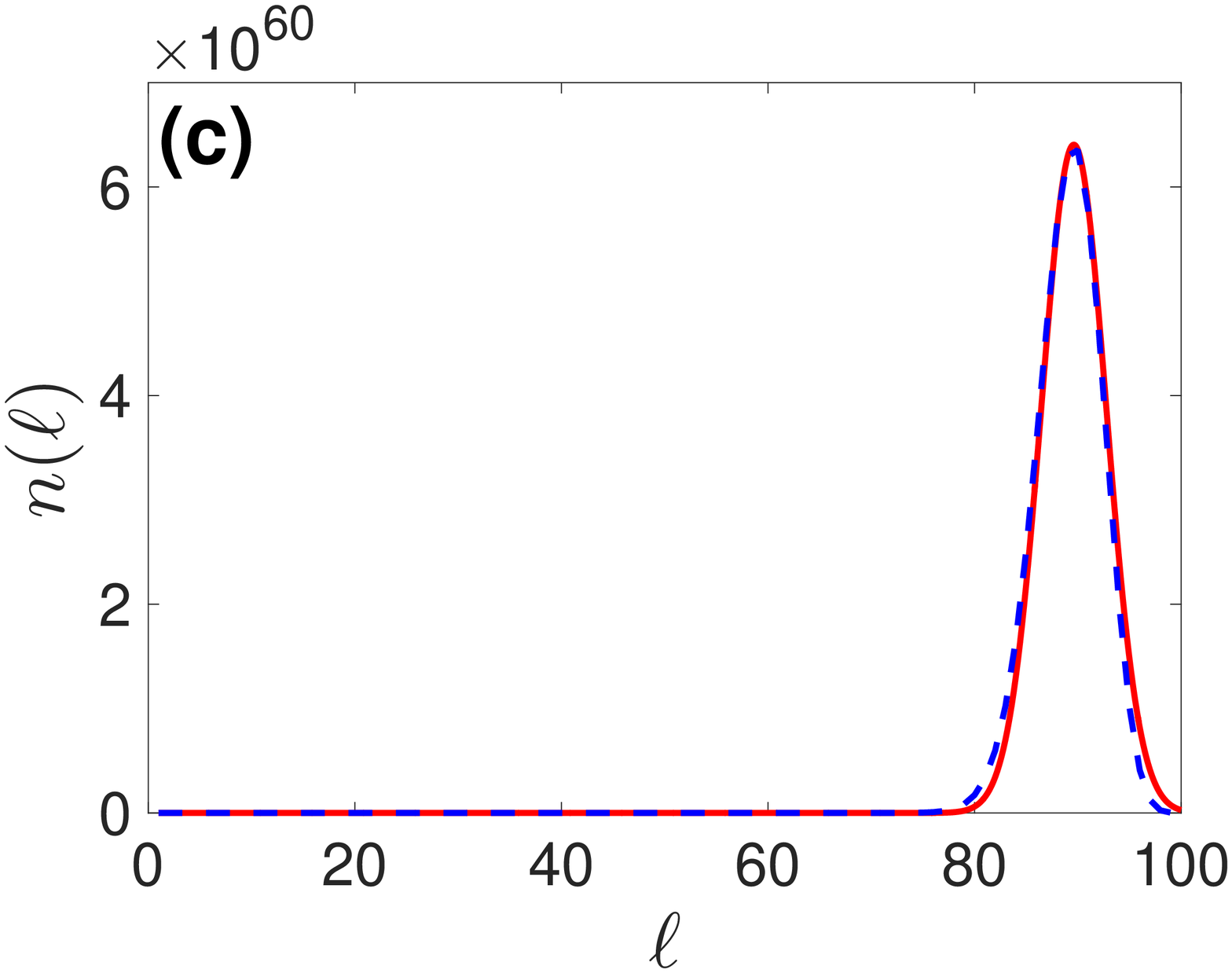}
}
\caption{
The number, 
$n(\ell)$, 
of SAW paths of length $\ell$, 
starting at a randomly chosen node, for a network 
of size $N=100$ and different values of $c$: 
$c=2$ (a), $c=5$ (b) and $c=7$ (c). 
The 
dashed lines 
(blue) 
are obtained from exact enumeration of the paths,
using 
Eq. (\ref{eq:n(ell)}).
The solid lines 
(red) are obtained from 
an asymptotic expression, 
namely the Gaussian approximation given by  
Eq. (\ref{eq:ell_approx}),
showing excellent agreement with the exact enumeration.
The function $n(\ell)$ exhibits a well defined and highly 
symmetric peak, which shifts to the right as $c$ is increased.
}
\label{fig:n(ell)}
\end{figure}

In Fig.
\ref{fig:n(ell)}
we present 
the number of SAW paths,
$n(\ell)$, obtained from Eq.
(\ref{eq:n(ell)})
for an ER network of size $N=100$
and three values of $c$
(dashed lines).
The function 
$n(\ell)$ has a well defined 
and highly symmetric peak,
which shifts to the right as $c$ is increased.
To obtain the location of the peak, we solve for 
the derivative 
$d n(\ell)/ d \ell = 0$, 
where 
$n(\ell)$ 
in taken from Eq.
(\ref{eq:n(ell)}).
We obtain 

\begin{equation}
\ell^{peak} \simeq N - \frac{1}{2 W \left[ \frac{c}{2 (N-1)}  \right]},
\end{equation}

\noindent
where $W(x)$ is the Lambert W function, also referred to as the
ProductLog function
\cite{Olver2010}.
In the limit of 
large and 
dilute networks, 
this expression can be 
simplified to 

\begin{equation}
\ell^{peak} \simeq N - \frac{N-1}{c} - \frac{1}{2}.
\label{eq:lpeak}
\end{equation}

\noindent
Expanding 
$\ln n(\ell)$ 
around
$\ell^{peak}$
to second order in $\ell$ 
leads to a Gaussian approximation of the form 

\begin{equation}
n(\ell) \simeq \frac{n^{tot}}{\sqrt{2 \pi \sigma^2}} 
e^{- \frac{\left(\ell - \ell^{peak}\right)^2}{2 \sigma^2}},
\label{eq:ell_approx}
\end{equation}

\noindent
where

\begin{equation}
\sigma^2 = \frac{1}{\psi^{(1)}\left( \frac{N-1}{c} + \frac{1}{2} \right) }, 
\label{eq:sigma}
\end{equation}

\noindent
and
$\psi^{(m)}(x)$
is the PolyGamma function of order $m$
\cite{Olver2010}.
For dilute networks, 
this converges to
$\sigma^2 = N/c$.
The prefactor, 
$n^{tot}$,
represents the total number of SAW paths of all possible lengths,
namely

\begin{equation}
n^{tot} = \sum_{\ell=1}^{N-1} n(\ell).
\label{eq:ntot0}
\end{equation}

\noindent
It is given by

\begin{equation}
n^{tot} = \frac{\sqrt{2 \pi \sigma^2} (N-1)!}
{\Gamma \left( \frac{N-1}{c} + \frac{1}{2} \right)}
\left( \frac{c}{N-1} \right)^{N - \frac{N-1}{c} - \frac{1}{2}}.
\label{eq:ntot1}
\end{equation}

\noindent
The Gaussian approximation of
Eq.
(\ref{eq:ell_approx}),
for the number of SAW paths, 
$n(\ell)$,
is shown in
Fig. 
\ref{fig:n(ell)},
for three values of $c$ (solid lines).
They are found to be in excellent agreement with the
exact results
(dashed lines).

The expressions presented above for $n(\ell)$ enumerate all the SAW paths
of length $\ell$ (starting from a random node $i$), 
regardless of whether they terminate after $\ell$ steps
or continue to form longer paths. 
To enumerate only the paths which terminate after $\ell$ steps, one needs
to multiply $n(\ell)$ by the termination probability, which is 
given by
$p_{\ell}(k=0) = \exp[-c(\ell)]$.
Therefore, the number of SAW paths which 
terminate after $\ell$ steps is

\begin{equation}
n_T(\ell) 
= n(\ell) e^{-c(\ell)}.
\label{eq:nT(ell)}
\end{equation}

\noindent
We find that the peak of this function is at

\begin{equation}
\ell_{T}^{peak} \simeq N - \frac{1}{2 W
\left[ \frac{c \exp(c/N)}{2 (N-1)}  \right]}.
\end{equation}

\noindent
For large and dilute networks, 
it can be approximated by

\begin{equation}
\ell_{T}^{peak} \simeq N - \frac{N-1}{c} e^{-c/N} - \frac{1}{2}.
\label{eq:lTpeak}
\end{equation}

\noindent
The function 
$n_T(\ell)$
can be approximated by a Gaussian
of the form

\begin{equation}
n_T(\ell) 
\simeq \frac{n_{T}^{tot}}{\sqrt{2 \pi \sigma_T^2}} 
e^{- \frac{\left(\ell - \ell_{T}^{peak}\right)^2}{2 \sigma_T^2}},
\label{eq:ellT_approx}
\end{equation}

\noindent
with

\begin{equation}
\sigma_T^2 = \frac{1}{\psi^{(1)}
\left( \frac{N-1}{c} e^{-c/N} + \frac{1}{2} \right) }. 
\label{eq:sigmaT}
\end{equation}

\noindent
The pre-factor

\begin{equation}
n_{T}^{tot} = \sum_{\ell=1}^{N-1} n_T(\ell) 
\label{eq:ntot2}
\end{equation}

\noindent
is given by

\begin{equation}
n_{T}^{tot} = 
\frac{\sqrt{2 \pi \sigma_T^2} (N-1)!}{\Gamma \left( \frac{N-1}{c} e^{-c/N} + \frac{1}{2} \right)}
\left( \frac{c}{N-1} \right)^{N - \frac{N-1}{c} e^{-c/N} - \frac{1}{2}}
e^{- \frac{c}{2N}-\exp(-c/N)}.
\label{eq:nTtot1}
\end{equation}

We now compare the results 
for the number of SAW paths which terminate
after $\ell$ steps, $n_T(\ell)$, vs. the total number, 
$n(\ell)$,
of paths of length $\ell$,
given by Eqs.
(\ref{eq:nT(ell)})
and
(\ref{eq:n(ell)}),
respectively.
Clearly, $n_T(\ell) \le n(\ell)$, for
$\ell=1,\dots,N-1$,
with equality at
$\ell = N-1$.
Since the termination probability increases with $\ell$,
the ratio 
$n_T(\ell)/n(\ell)$ 
is a monotonically 
increasing function.
As a result,
the peak of $n_T(\ell)$
is shifted to the right
with respect to the peak
of $n(\ell)$,
namely
$\ell_T^{peak} > \ell^{peak}$.
This can be easily confirmed by 
a comparison between
Eqs.
(\ref{eq:lpeak})
and
(\ref{eq:lTpeak}),
using the fact that
$W(x)$ is a
monotonically increasing function 
for $x>0$.

In case of a dilute network,
both functions can be approximated by
Gaussian forms,
given by Eqs.
(\ref{eq:ellT_approx}) 
and
(\ref{eq:ell_approx}),
respectively.
Comparing between Eqs.
(\ref{eq:sigma})
and
(\ref{eq:sigmaT}),
and using the fact that 
$\psi^{(1)}(x)$ 
is a monotonically decreasing function for $x>0$,
one can show that the peak of $n_T(\ell)$ is narrower than
the peak of $n(\ell)$,
namely
$\sigma_T < \sigma$.

\section{The distribution of SAW path lengths}

Consider an SAW on an ER network, 
which starts from a node with degree
$k \ge 1$ (non-isolated node).
The SAW hops between nearest neighbor nodes until it reaches a node
whose all neighbors have already been visited. 
At that stage the SAW has no exit link and it is terminated.
In the complementary picture of a pruned network, the walker hops until
it reaches a node which becomes isolated upon its arrival.
The probability that a node is isolated at time $t$ is
$p_t(k=0)$. 
Therefore, the probability that the SAW will proceed from time
$t$ to time $t+1$ 
is 
given by the conditional probability
$P(d > t|d > t-1) = 1-p_t(k=0)$,
where the random variable $d$ denotes the distance 
pursued along
the SAW path. 
Thus, the probability that the path length of the SAW will
be longer than $\ell$ is given by

\begin{equation} 
P(d>\ell) = P(d>0) \prod_{t=1}^{\ell} P(d > t|d > t-1),
\label{eq:cond}
\end{equation}

\noindent
where $P(d>0)=1$
(since the initial node is not isolated).
Thus, this tail distribution takes the form

\begin{equation} 
P(d>\ell) = \prod_{t=1}^{\ell} \left[1-p_{t}(k=0)\right].
\label{eq:cond2}
\end{equation}

\noindent
While Eq.
(\ref{eq:cond2})
applies to any network, 
in the case of ER networks we have an explicit expression for
the probability of a node to become isolated,
namely 
$p_t(k=0)=\exp[-c(t)]$.
Therefore, the tail distribution takes the form

\begin{equation}
P(d>\ell)
=\prod_{t=1}^{\ell}\left[1-e^{-c(t)}\right].
\label{eq:cond3}
\end{equation}

\begin{figure}
\centerline{
\includegraphics[width=8cm]{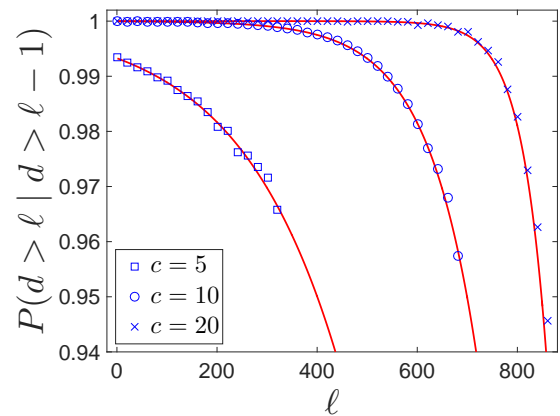}
}
\caption{
The conditional probability 
$P(d > \ell|d > \ell-1)$ 
vs. $\ell$,
obtained from Eq.
(\ref{eq:cond_c(t)})
(solid lines)
and 
from numerical simulations of SAWs 
(symbols)
on ER networks of
%coming from a simulation of a node-deleting random 
%walk on a network of 
size $N=1000$ and initial 
connectivities $c=5$, $10$ and $20$ 
(squares, circles and crosses, respectively).
The analytical and numerical results are found to
be in good agreement.
} 
\label{fig:P_cond}
\end{figure}

\noindent
The validity of this expression relies on the validity 
of the equation 

\begin{equation}
P(d > \ell|d > \ell-1)=1-e^{-\left(1- \frac{\ell}{N-1} \right)c} 
\label{eq:cond_c(t)}
\end{equation}

\noindent
for an ER network. 
In Fig. 
\ref{fig:P_cond} 
we present
the conditional probability 
$P(d > \ell|d > \ell-1)$ 
vs. $\ell$ for a network 
of size $N=1000$ and for three values of $c$. 
The analytical results (solid lines) obtained from Eq.
(\ref{eq:cond_c(t)})
are found to be in good agreement with numerical simulations
(symbols),
confirming the validity of this equation.
Note that the numerical results become more noisy as $\ell$
increases, due to diminishing statistics, 
and eventually terminate.
This is particularly apparent for the smaller values of $c$.

To obtain a closed form expression for the tail distribution, 
$P(d>\ell)$,
we
take the natural 
logarithm on both sides of Eq.
(\ref{eq:cond3}).
This leads to

\begin{equation}
\ln \left[P\left(d>\ell\right)\right] =
\sum_{t=1}^{\ell} \ln \left[1-\exp\left(\frac{ct}{N-1}-c\right) \right].
\end{equation}

\noindent
Approximating this sum by an integral we obtain

\begin{equation}
\ln \left[P\left(d>\ell\right)\right]
\simeq
\int_{1/2}^{\ell+1/2}\ln \left[1-\exp \left( \frac{ct}{N-1}-c \right) \right]dt.
\end{equation}

\noindent
This integral is in fact a partial Bose-Einstein integral, which 
can be expressed in terms of
the Polylogarithm $Li_{n}(x)$
function
\cite{Olver2010}

\begin{equation}
P(d>\ell)
\simeq
\exp \left\{\frac{N-1}{c}\left[Li_{2}
\left(e^{- \left(1- \frac{1}{2(N-1)}\right) c}
\right)-Li_{2}\left(e^{- \left(1- \frac{\ell+1/2}{N-1} \right) c}
\right)\right]\right\}.
\label{eq:P(d>l)}
\end{equation}

\noindent
The probability distribution
$P(\ell)$ 
is obtained from the
tail distribution by

\begin{equation}
P(\ell) = P(d > \ell-1) - P(d>\ell).
\label{eq:diff0}
\end{equation}

\noindent
In the analysis below, we replace the difference 
in Eq.
(\ref{eq:diff0})
by a derivative.
This replacement is justified either for very smooth
fucntions or for large values of $\ell$. 
Indeed, the function $P(d>\ell)$ satisifies these conditions
for both sparse and dense networks.
For small values of $c$, it is smooths over its entire range,
while for larger values of $c$ it exhibits a sharp variation only 
in the range of large $\ell$.

Replacing the difference by a derivative we obtain

\begin{equation}
P(\ell)=-\frac{d P(d>\ell)}{d \ell}.
\label{eq:diff}
\end{equation}

\noindent
Plugging Eq.
(\ref{eq:P(d>l)})
into Eq.
(\ref{eq:diff})
we obtain

\begin{equation}
P(\ell)
\simeq
-\ln\left[1-e^{- \left(1- \frac{\ell+1/2}{N-1}\right) c}\right]
\cdot P\left(d>\ell\right).
\label{eq:P(l)}
\end{equation}

\begin{figure}
\centerline{
\includegraphics[width=16cm]{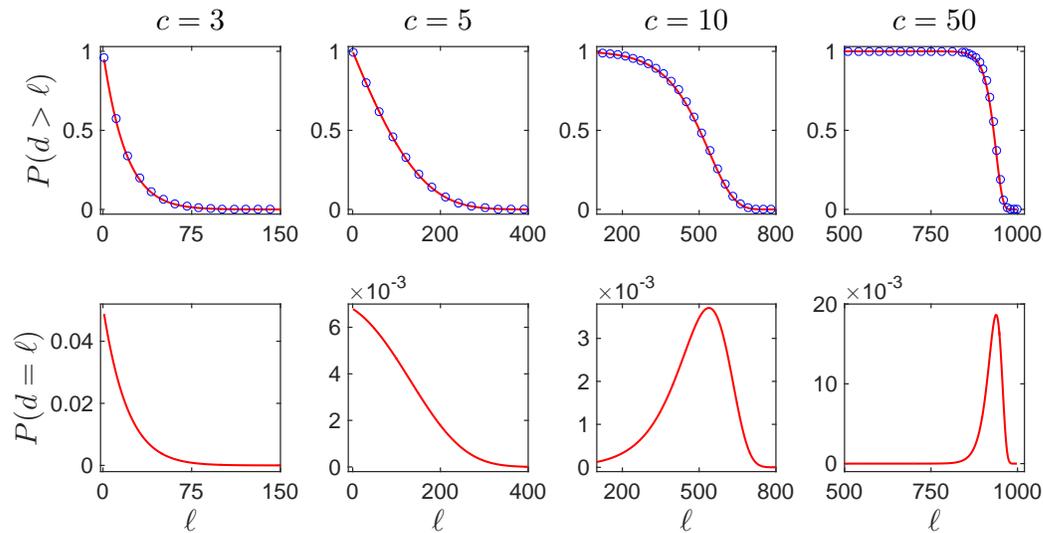}
}
\caption{
The distributions 
of path lengths of SAWs on ER networks
of size $N=1000$ 
and 
$c=3$, $5$, $10$ and $50$.
The tail distributions 
$P(d > \ell)$,
obtained from Eq.
(\ref{eq:P(d>l)}) (solid lines)
and from numerical simulations (circles) 
are presented in the top row,
with excellent agreement between the two.
The corresponding probability density functions 
$P(\ell)$,
obtained from Eq.
(\ref{eq:P(l)}) 
are shown in the bottom row.
The agreement with the numerical results is already established
in the top row and therefore the numerical data is not shown 
in the bottom row.
}
\label{fig:P(d>ell)}
\end{figure}

\noindent
For large networks ($N \gg 1$)
one can further approximate
Eqs.
(\ref{eq:P(d>l)})
and
(\ref{eq:P(l)})
by

\begin{equation}
P\left(d>\ell\right)
\simeq
\exp\left[-\frac{N}{c}e^{-c}
\left(e^{\frac{c}{N}\ell}-1\right)\right]
\label{eq:tail2}
\end{equation}

\noindent
and

\begin{equation}
P(\ell)
\simeq
\exp\left[-\frac{N}{c}e^{-c}
\left(e^{\frac{c}{N}\ell}-1\right)
-\left(1-\frac{\ell+1}{N}\right)c
\right],
\label{eq:pdf2}
\end{equation}

\noindent
respectively.
In Fig. 
\ref{fig:P(d>ell)} 
we present the distributions 
of path lengths of SAWs on ER networks
of size $N=1000$, for different values of $c$.
The tail distributions, 
$P(d > \ell)$, 
are shown in the top row and
the corresponding probability density functions, 
$P(\ell)$, 
are shown in the bottom row.
The analytical results (solid lines), 
obtained from 
Eqs.
(\ref{eq:P(d>l)})
and
(\ref{eq:P(l)}),
are found to be in excellent agreement with numerical 
simulations (circles).
In fact, the approximated 
expressions of Eqs.
(\ref{eq:tail2})
and
(\ref{eq:pdf2})
provide results which are practically indistinguishable
from the more accurate expressions presented in 
Fig.
\ref{fig:P(d>ell)}. 
In the numerical simulations, the initial node of the SAW is chosen randomly
among the nodes on the largest connected cluster.
This is justified because for $c \ge 3$ less than one percent
of the nodes reside on small isolated clusters. 
Fig. \ref{fig:P(d>ell)} reveals three different qualitative behaviors of 
$P(\ell)$.
For small values of $c$ (sparse networks), 
$P(\ell)$ 
is a monotonically
decreasing function. 
As $c$ is increased, 
$P(\ell)$ forms a peak and 
becomes broader and more symmetric. 
In the limit of dense 
networks the peak becomes narrower as it shifts to the right.
As $c/(N-1) \rightarrow 1$, 
it approaches a delta function at 
$\ell=N-1$. 
Further insight 
about $P(\ell)$
is given below 
in the context of
central and dispersion measures.

It is interesting to note that the expressions for the 
distribution of SAW path lengths for large networks, presented in Eqs.
(\ref{eq:tail2})
and
(\ref{eq:pdf2}),
coincide with the corresponding equations of the 
Gompertz distribution 
\cite{Johnson1995}.
In particular,
the tail distribution of the Gompertz distribution 
for a random variable $X$ takes the form

\begin{equation}
P(X>x) =
\exp\left[-\eta
\left(e^{bx}-1\right)\right] 
\label{eq:Gompertz}
\end{equation}

\noindent
for $x \ge 0$.
Inserting the scale parameter

\begin{equation}
b = \frac{c}{N},
\label{eq:scale}
\end{equation}

\noindent
and the shape parameter 

\begin{equation}
\eta=\frac{N}{c}e^{-c}
\label{eq:shape}
\end{equation}

\noindent
into Eq.
(\ref{eq:Gompertz})
gives rise to Eq.
(\ref{eq:tail2}).

The Gompertz law describes the distribution of adult lifespans
\cite{Gompertz1825,Shklovskii2005} 
as well as 
various other survival probabilities,
such as the failure rates of computer codes 
\cite{Ohishi2008}.
The very old observation, attributed to 
Halley 
\cite{Halley1693} 
and Euler 
\cite{Euler1760}, 
is that an exponential life 
expectancy of the form 
$S(t)=\exp(-t/t_0)$,
where $S(t)$ is the survival probability of an individual, 
and $t_0$ being a characteristic life span (say $70$), 
would entail millions of people with the age of $200$. 
Gompertz suggested that the mortality rate is not a constant,
as implied by the exponential law, but rather 
increases exponentially with age, 
which explains why the longest recorded 
human life did not exceed $123$ years
\cite{Robine1998}. 

In our case, this observation provides an interesting narrative 
for the life expectancy of an SAW on the network - the termination rate 
increases exponentially with time as a result of the fact 
that the SAW prunes the network along its walk and makes it sparser.
There are however two important differences
between the Gompertz distribution 
$P(X>x)$ of Eq. 
(\ref{eq:Gompertz})
and the tail distribution
$P(d>\ell)$
of Eq.
(\ref{eq:tail2}). 
The first difference is that
$P(d>\ell)$  
describes a discrete distribution 
over integer values of $\ell$
while the Gompertz distribution describes a continuous random variable.
The second difference is that
the Gompertz law is unbounded (valid for any $x \ge 0$), while 
the longest possible SAW path on a network 
is of length $\ell=N-1$.
The first difference is important in the limit of sparse networks,
where the SAW path lengths are small and the discrete nature of $\ell$
is apparent.
The second difference is important in the limit of dense networks,
where the SAW path lengths approach their maximal value.

\section{Central measures of the SAW path length distribution}

In order to characterize the distribution of path lengths of the SAW we
derive expressions for the mean, median and mode of this distribution.
The mean of the distribution can be obtained
from the tail-sum formula

\begin{equation}
\ell_{mean}(N,c) =
\sum_{\ell=0}^{N-2} P(d>\ell). 
\label{eq:tailsum1}
\end{equation}

\noindent
Assuming that the initial node is not isolated,
this sum can be written in the form

\begin{equation}
\ell_{mean}(N,c) =
1 + \sum_{\ell=1}^{N-2} P(d>\ell). 
\label{eq:tailsum2}
\end{equation}

\noindent
Expressing the sum 
as an integral we obtain

\begin{equation}
\ell_{mean}(N,c) =
1 + \int_{1/2}^{N-3/2} P(d>\ell)d\ell.
\end{equation}

\noindent
Plugging in $P(d>\ell)$ 
from Eq.
(\ref{eq:tail2}),
the resulting integral has the closed form 
expression 

\begin{equation}
\ell_{mean}(N,c)
\simeq
1+\frac{N}{c}
\left[Ei\left(-\frac{N}{c}e^{-\frac{3c}{2N}}\right)
-Ei\left(-\frac{N}{c}e^{-\left( 1- \frac{1}{2N}\right)c}\right)\right]
\exp \left( \frac{N}{c}e^{-c} \right), 
\label{eq:mean1}
\end{equation}

\noindent
where $Ei(x)$ is the
exponential integral 
\cite{Olver2010}.
In the limit of large $N$ one can write an 
approximated expression using only elementary functions,
by expanding in the parameter
$\eta = (N/c) e^{-c}$,
resulting in two regimes.
In sparse networks, 
where $\eta > 1$,
the mean path length is given by

\begin{equation}
\ell_{mean}(N,c) \simeq 1 + e^c.
\label{eq:mean2_0} 
\end{equation}

\noindent
In dense networks,
where $\eta < 1$,
the mean path length is

\begin{equation}
\ell_{mean}(N,c)
\simeq
1+
\left[ N - \frac{N}{c} \left(\ln \frac{N}{c} +\gamma \right)
+ \left( \frac{N}{c} \right)^2 e^{-c}
- \frac{1}{4} \left( \frac{N}{c} \right)^3 e^{-2c}
\right]
\exp \left( \frac{N}{c}e^{-c} \right),
\label{eq:mean2}
\end{equation}

\noindent
where $\gamma$ is the Euler-Mascheroni constant
\cite{Olver2010}.

The median 
of $P(\ell)$
is obtained 
by equating 
the right hand side of
Eq.
(\ref{eq:tail2})
to 
$1/2$
and solving for $\ell$.
The resulting expression is

\begin{equation}
\ell_{median}(N,c) \simeq
\frac{N}{c} \ln \left(1+e^{c}\cdot\frac{c}{N}\ln2\right).
\label{eq:median1}
\end{equation}

\noindent
The mode of 
the distribution of path lengths
is the value of $\ell$ which maximizes
$P(\ell)$. 
For $c < \ln N$ the distribution is
monotonically decreasing and the maximum is obtained for
$\ell=1$.
For larger values of $c$, the distribution develops a peak,
where the derivative of $P(\ell)$ vanishes. 
Using 
$d P(\ell)/d \ell = 0$ 
in Eq.
(\ref{eq:pdf2})
we obtain

\begin{equation}
\ell_{mode}(N,c) \simeq 
\left\{
\begin{array}{ll}
1 & \text{   if  } c \le c_0 
\\
\left \lfloor N - \frac{N}{c} \ln \left(\frac{N}{c} \right) 
\right \rfloor & 
\text{   if  } c > c_0,
\end{array}
\right.
\label{eq:mode1}
\end{equation}

\noindent
where $\lfloor x \rfloor$
is the integer part of $x$.
The transitional $c_{0}$ between
these two regimes of the probability density function 
is obtained by equating 
$\ell_{mode}$
to $1$ and solving for $c$, 
given approximately by 
$\ln c+c = \ln N$.
Solving this equation we find that

\begin{equation}
c_0 = W(N),
\end{equation}

\noindent
where 
$W(x)$ 
is
the Lambert W function
\cite{Olver2010}. 
Note that $c_0$ is also the point at which
the shape parameter $\eta$ of the Gompertz distribution,
given by Eq. (\ref{eq:shape}),
which also appears in the discussion that follows 
Eq. (\ref{eq:mean1}),
is equal to $1$. 
It separates the small $c$ regime, where $\eta>1$,
from the large $c$ regime, where $\eta<1$.

Interestingly, in the regime
$c \gg c_0$
we find that all the three central measures presented above
converge to the same asymptotic expression given by

\begin{equation}
\ell_{m}(N,c) \simeq
N \left[1-\frac{1}{c}\ln\left(\frac{N}{c}\right)\right].
\label{eq:mmm}
\end{equation}

\noindent
This means that the path lengths of typical SAWs converge towards $N$
as $c$ is increased. 
Thus, a typical SAW path becomes a Hamiltonian path
as $c \rightarrow N$.

\begin{figure}
\centerline{
\includegraphics[width=5cm]{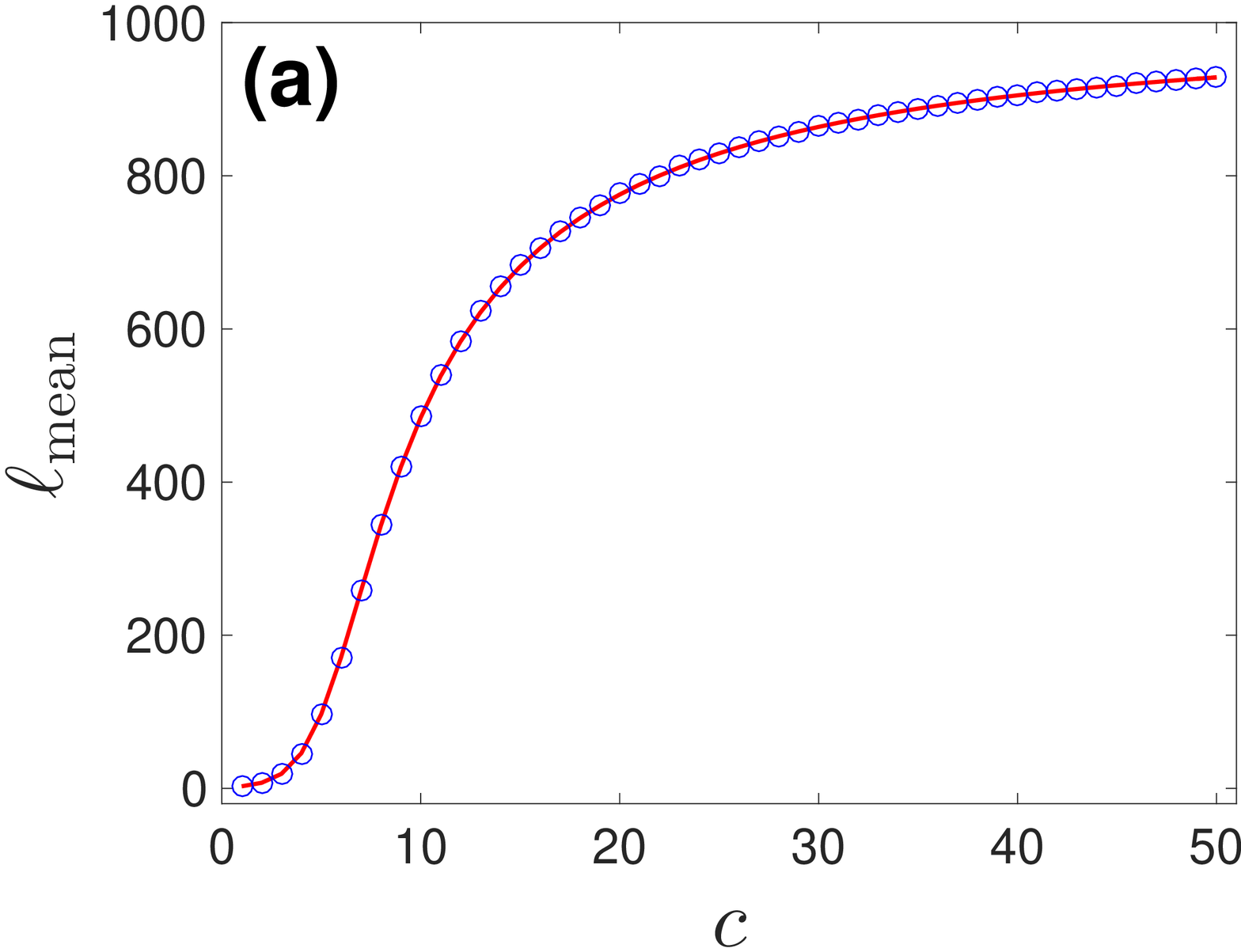}
\includegraphics[width=5cm]{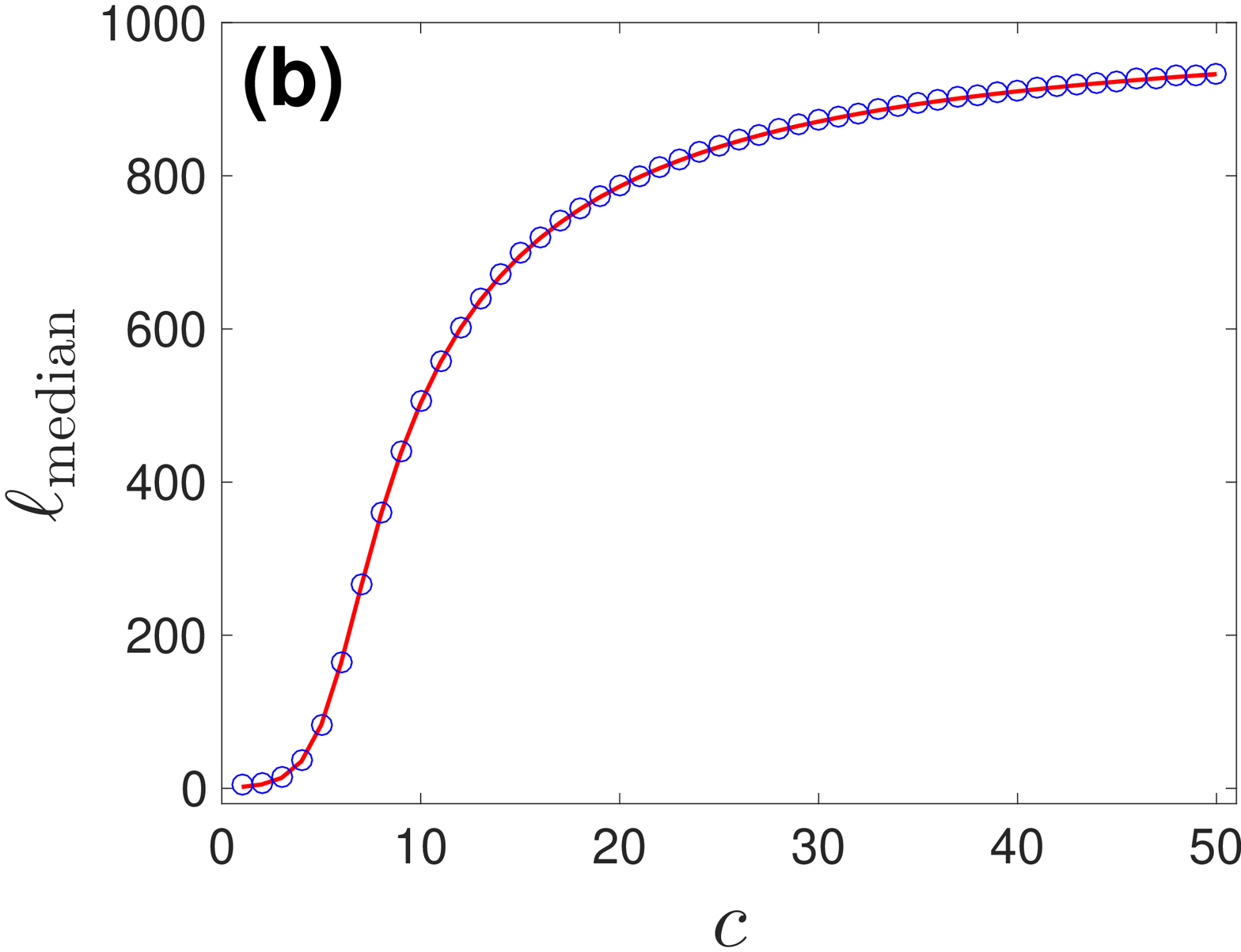}
\includegraphics[width=5cm]{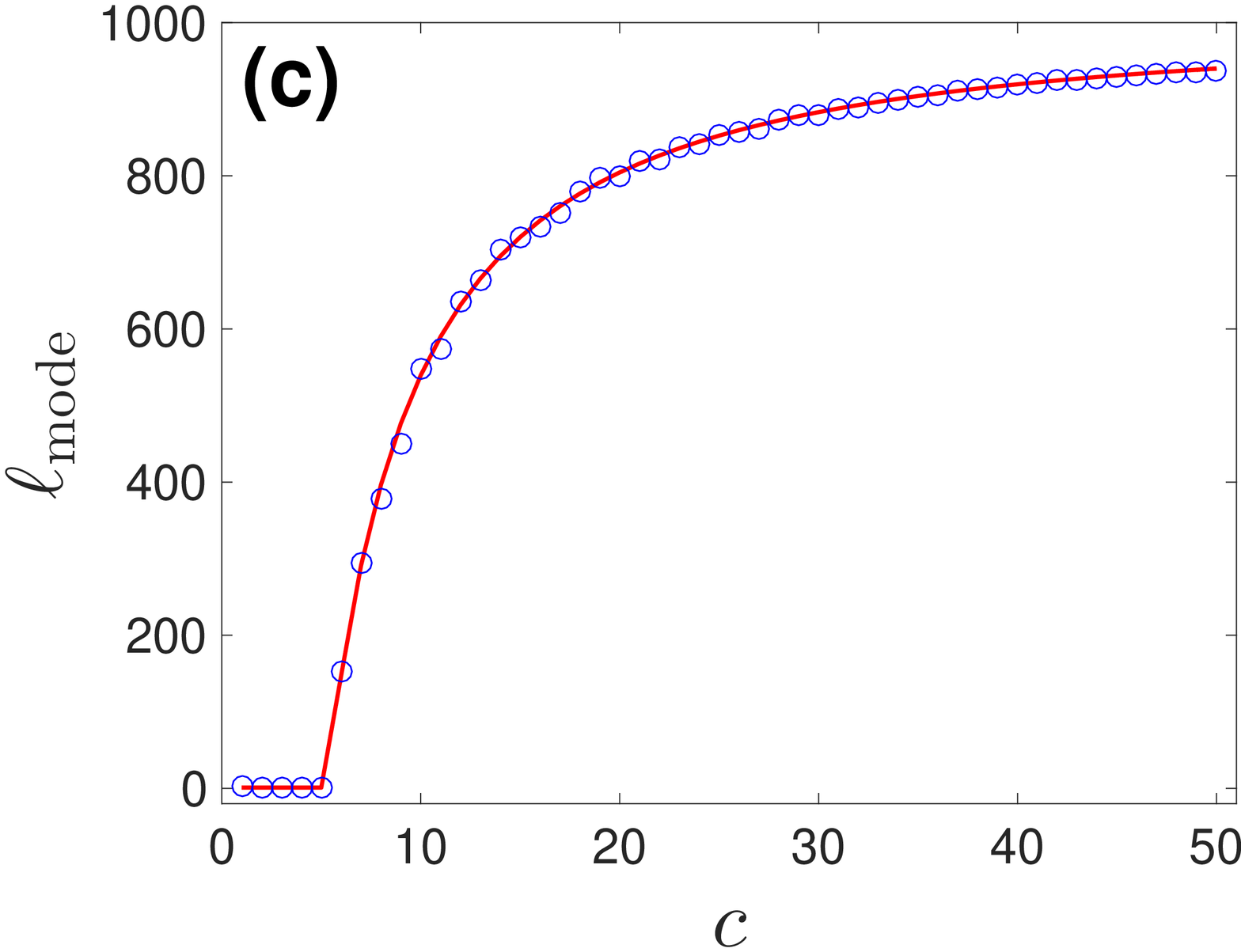}
}
\caption{
The mean (a), median (b) and mode (c) of the distribution of path
lengths of SAWs vs. the initial connectivity, $c$ for
ER networks of size
$N=1000$.
The analytical results (solid lines) for the mean, median and mode are
obtained from Eqs.
(\ref{eq:mean1}),
(\ref{eq:median1}),
and
(\ref{eq:mode1}),
respectively.
The results are in excellent agreement with numerical 
simulations 
(circles). 
}
\label{fig:mmm}
\end{figure}

In Fig.  
\ref{fig:mmm}
we present
the central measures 
$\ell_{mean}(N,c)$ (a)
$\ell_{median}(N,c)$ (b)
and
$\ell_{mode}(N,c)$ (c),
vs. $c$,
for ER networks of size $N=1000$.
The solid lines represent the analytical results,
obtained from Eqs.
(\ref{eq:mean1}),
(\ref{eq:median1}),
and
(\ref{eq:mode1}),
respectively.
They are found to be in excellent agreement with numerical simulations
(circles).

\section{Measures of dispersion of the SAW path length distribution}

The width of the path length distribution 
can be characterized by the
variance
$\sigma_{\ell}^2 = \langle \ell^2 \rangle - \langle \ell \rangle^2$,
where
$\langle \ell^n \rangle$,
is given by
the tail-sum formula
\cite{Pitman1993}

\begin{equation}
\langle \ell^n \rangle = 
\sum_{\ell=0}^{N-2} [(\ell+1)^n - \ell^n] P(d>\ell).
\label{eq:tail_sum}
\end{equation}

\noindent
Neglecting exponential corrections we can set the upper limit of the 
sum to $\infty$ and replace the sum by an integral. 
Plugging in 
$P(d>\ell)$ 
from Eq.
(\ref{eq:tail2})
results in the following closed-form expression for the variance

\begin{equation}
\sigma _\ell ^2 = \frac{{{N^2}}}{{{c^2}}}{e^\eta }
\left[ {{\gamma ^2} + \frac{{{\pi ^2}}}{6} 
- {e^\eta }[Ei{{( { - \eta } )}]^2} 
- 2\eta  \cdot {}_3{F_3}\left( {\begin{array}{*{20}{c}}
{1,1,1}\\
{2,2,2}
\end{array}, - \eta } \right) + 2\gamma 
\ln \eta  + {{\left( {\ln \eta } \right)}^2}} \right] , 
\label{eq:STD}
\end{equation}

\noindent
where 
$\eta=\frac{N}{c}{e^{-c}}$ 
is the shape parameter of the corresponding 
Gompertz distribution,
given by Eq. (\ref{eq:shape}), 
and ${}_3{F_3}$ is a generalized hypergeometric function 
\cite{Olver2010}.
It can be expressed by

\begin{equation}
{}_3{F_3}\left( {\begin{array}{*{20}{c}}
{1,1,1}\\
{2,2,2}
\end{array},x} \right) 
= \sum\limits_{k = 0}^\infty  
{\frac{{{x^k}}}{{{{\left( {k + 1} \right)}^3}k!}}}.
\label{eq:3F3}
\end{equation}

\noindent
For small values of $c$,
namely $c < c_0$, 
expanding the right hand side of Eq.
(\ref{eq:STD})
in powers of 
the small parameter
$1/\eta$
we obtain

\begin{equation}
\sigma _\ell ^2 \simeq {e^{2c}}\left[ {1 - \frac{4}{\eta} + 
\frac{17}{\eta^2} 
+ \mathcal O\left( 
\frac{1}{\eta^3}
\right)} \right].
\label{eq:STD-small}
\end{equation}

\noindent
For large values of $c$,
namely $c > c_0$, 
we use $\eta$ as a small parameter to
express the 
right hand side of Eq.
(\ref{eq:STD})
as

\begin{equation}
\sigma _\ell ^2 
\simeq 
\frac{\pi ^2}{6} \frac{N^2}{c^2}
\left[ {1 + 
\eta
\left( {\frac{{{\pi ^2} - 12 + 12\gamma  
- 6{\gamma ^2}}}{{{\pi ^2}}} + 
12\left( {1 - \gamma } \right)
\ln \eta
- 6 (\ln \eta)^2 } \right)
+ \mathcal O(\eta^2)  } \right] .
\label{eq:STD-large}
\end{equation}

Another way to
characterize the width of the path length distribution 
is to calculate the
interquartile range (IQR). 
It is defined as 
$IQR = \ell_{3/4} - \ell_{1/4}$,
where
$\ell_{3/4}$
is the upper quartile
[namely, $P(d \le \ell_{3/4}) = 3/4$]
and
$\ell_{1/4}$
is the lower quartile
[namely, $P(d \le \ell_{1/4}) = 1/4$]. 
The upper (lower) quartile is
obtained by equating 
Eq.
(\ref{eq:tail2})
to 
$1/4$ ($3/4$)
and solving for 
$\ell_{3/4}$
($\ell_{1/4}$).
The result is

\begin{equation}
IQR(N,c)
\simeq
\frac{N}{c} \ln 
\left(1 + \frac{ \ln 3}{\frac{N}{c}e^{-c} +  \ln \frac{4}{3}} \right).
\label{eq:IQR}
\end{equation}

\begin{figure}
\centerline{
\includegraphics[width=7cm]{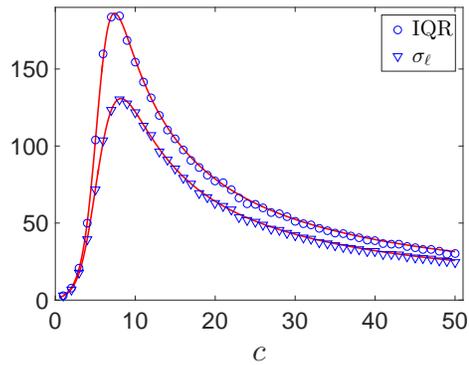}
}
\caption{
The standard deviation $\sigma_\ell$ (triangles) and the 
inter quartile interval, IQR (circles),
for the distribution of path 
lengths of SAWs as a function of  the initial connectivity, $c$, for
ER networks of size
$N=1000$.
The analytical results (solid lines), 
obtained from Eqs.
(\ref{eq:STD}) and 
(\ref{eq:IQR}) respectively,
are in excellent agreement with numerical 
simulations (symbols). 
}
\label{fig:sigma}
\end{figure}

In Fig. 
\ref{fig:sigma} 
we present 
$\sigma_\ell$ 
and 
$IQR$ 
as a function of $c$. 
The analytical results for both measures 
(solid lines)
are found to be in excellent agreement with 
the numerical results
(symbols).
Both measures indicate a maximal dispersion  
around $c=c_0$, which marks the crossover
between the regimes of sparse and dense networks. 

\section{Extreme value statistics of the SAW path length}

Another way to characterize the dispersion of the distribution is to
express it in terms of extreme value statistics.
To this end, we examine the lengths 
$\ell_{max}$
and
$\ell_{min}$
of the longest and shortest paths
among $r$
independent realizations of the SAW on an $ER[N,c/(N-1)]$ network.
The expected value of
$\ell_{max}$
is given by the condition
$P(d > \ell) < 1/r$
while the expected value of
$\ell_{min}$
is given by
$P(d < \ell) < 1/r$.

Using the expression for 
$P(d>\ell)$, 
given by Eq.
(\ref{eq:tail2}),
we obtain

\begin{equation}
\ell_{max} = \frac{N}{c} \ln \left( 1 + \frac{c}{N} e^c \ln r \right)
\label{eq:l_max}
\end{equation}

\noindent
and

\begin{equation}
\ell_{min} = \frac{N}{c} \ln 
\left[ 1 + \frac{c}{N} e^c \ln \left( \frac{r}{r-1} \right) \right].
\label{eq:l_min}
\end{equation}

\noindent
In the limit of a large number of realizations,
$r \gg 1$,
the expression for $\ell_{min}$ can be simplified to

\begin{equation}
\ell_{min} \simeq \frac{N}{c} \ln 
\left( 1 + \frac{c}{Nr} e^c   \right).
\label{eq:l_min2}
\end{equation}

\begin{figure}
\centerline{
\includegraphics[width=7cm]{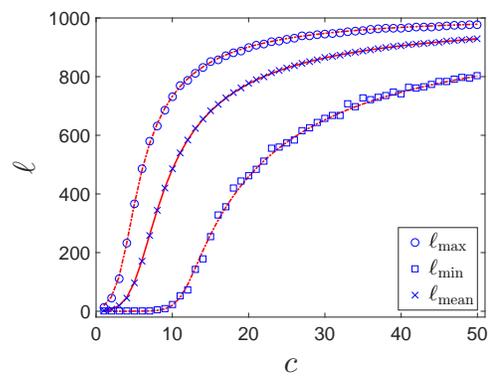}
}
\caption{
The lengths, 
$\ell_{max}$
and
$\ell_{min}$
of the longest (circles) and shortest (squares) SAW paths,
respectively,
as a function of $c$,
among $r=10^4$
independent SAW realizations 
on an ER network of size $N=1000$.
The solid lines are obtained from Eq.
(\ref{eq:l_max})
for the longest path and from Eq.
(\ref{eq:l_min})
for the shortest path, and both are in excellent agreement
with the numerical simulations. 
The average lengths (crosses) 
are shown for comparison. 
}
\label{fig:ell(c)}
\end{figure}

In Fig.
\ref{fig:ell(c)}
we present the lengths 
$\ell_{max}$
and
$\ell_{min}$
of the longest and shortest paths 
among 
$r=10^4$ 
independent SAW realizations 
vs. the initial connectivity, $c$,
on an ER network of size $N=1000$.
The theoretical results for the 
longest paths (solid line),
obtained from Eq.
(\ref{eq:l_max}),
are in excellent agreement with the numerical simulations
(circles).
The theoretical results for the
shortest paths (solid line),
obtained from Eq.
(\ref{eq:l_min}),
are also in excellent agreement with the numerical simulations
(squares).
The average lengths 
$\ell_{mean}$,
obtained from the simulations
(crosses) 
and from Eq. 
(\ref{eq:mean2})
(solid line),
are shown for comparison. 

\begin{figure}
\centerline{
\includegraphics[width=7cm]{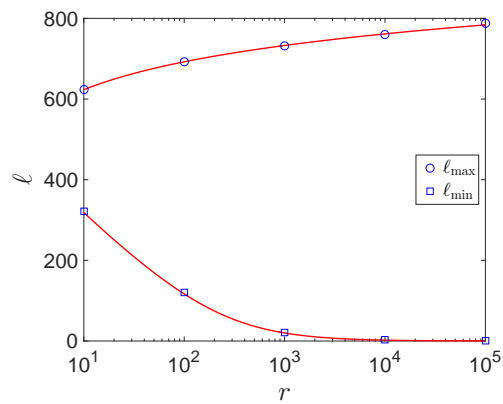}
}
\caption{
The lengths, 
$\ell_{max}$
and
$\ell_{min}$
of the longest (circles) and shortest (squares) SAW paths,
respectively,
as a function of the number of independent realizations, $r$, 
of the SAW on an ER network with
$N=1000$ and $c=10$.
The solid lines are obtained from Eq.
(\ref{eq:l_max})
for the longest path and from Eq.
(\ref{eq:l_min})
for the shortest path, and both are in excellent agreement
with the numerical simulations. 
}
\label{fig:ell(r)}
\end{figure}

In Fig.
\ref{fig:ell(r)}
we present the lengths 
$\ell_{max}$
and
$\ell_{min}$
of the longest and shortest paths 
as a function of the number of independent realizations, $r$, 
of the SAW on an ER network with
$N=1000$ and $c=10$.
The theoretical results for the longest paths (solid line), 
obtained from Eq.
(\ref{eq:l_max}),
are in excellent agreement with numerical simulations (circles).
The theoretical results for the shortest paths (solid line),
obtained from Eq.
(\ref{eq:l_min}),
are also in excellent agreement with
the numerical simulations (squares). 

\section{Summary and Discussion}

We have studied SAW paths on finite ER networks.
In practice, SAWs on networks delete the nodes they visit, 
thus gradually reducing the network size.
We have shown that the pruned network maintains its ER character,
with a linearly decreasing average degree, $c(t)$.
We obtained an exact formula for the number of SAW paths,
$n_T(\ell)$, which terminate after $\ell$ steps and analyzed
its behavior as a function of the initial connectivity, $c$.
We also studied the disribution of path lengths $P(\ell)$ for
SAW paths which are actually 
pursued by a self-avoiding random walker
starting from a random initial
node, $i$. We obtained an analytical expression for $P(\ell)$ as
well as a large $N$ approximation valid for large networks.
It was found that for low initial connectivity, $P(\ell)$ is a
monotonically decreasing function, while for larger values of
$c$ it exhibits a well rounded peak, which shifts to the right
as $c$ is increased.
To characterize the distribution 
$P(\ell)$ 
we obtained 
analytical expressions for its 
central measures, namely the mean, median and mode.
We also derived measures for the dispersion of $P(\ell)$,
namely the standard deviation and the inter-quartile range.
Studying the extreme value statistics of 
the SAW path lengths,
we obtained
the expectation value of the longest and shortest paths among
$r$ independent SAW paths.

\begin{figure}
\centerline{
\includegraphics[width=7cm]{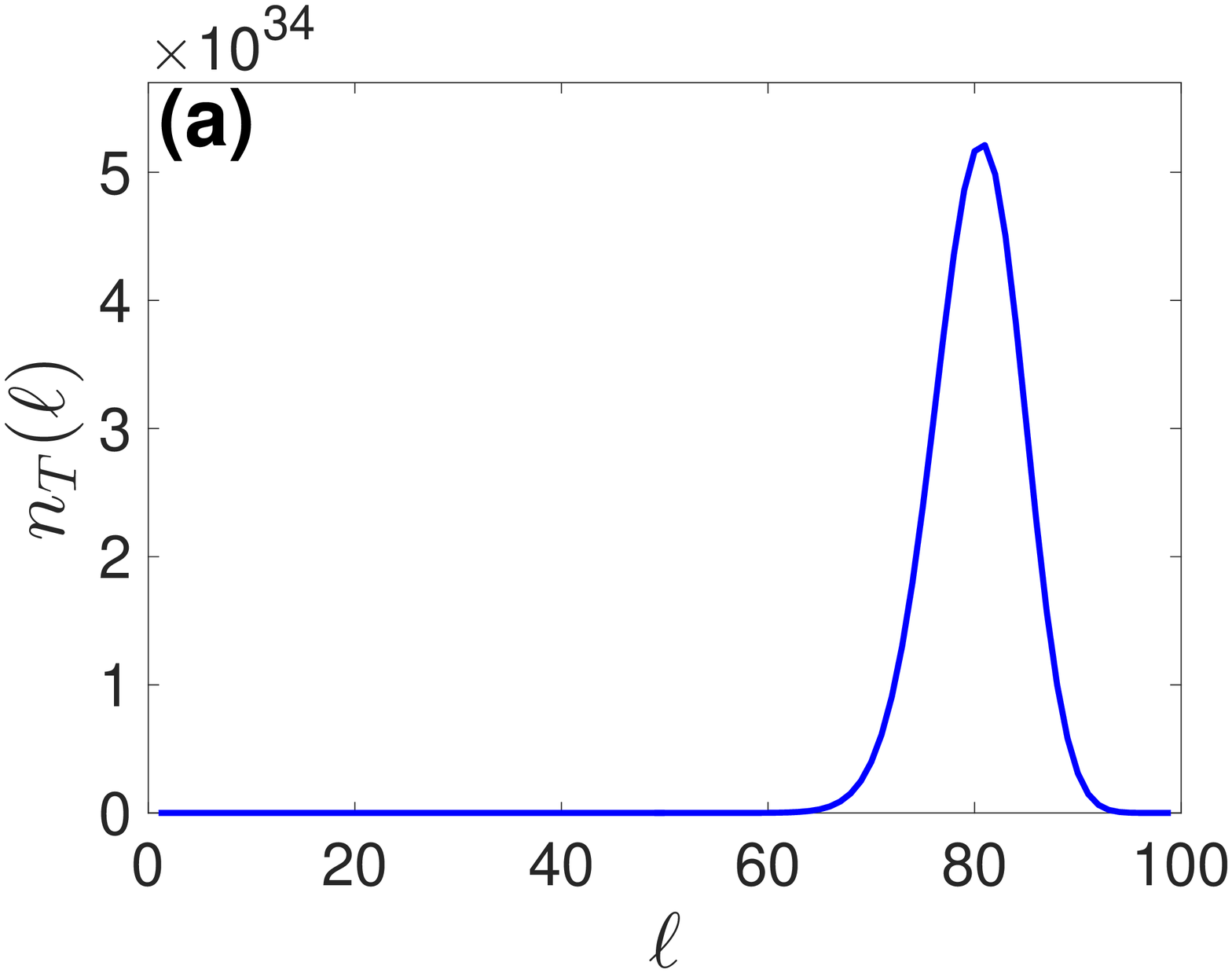}
\includegraphics[width=7cm]{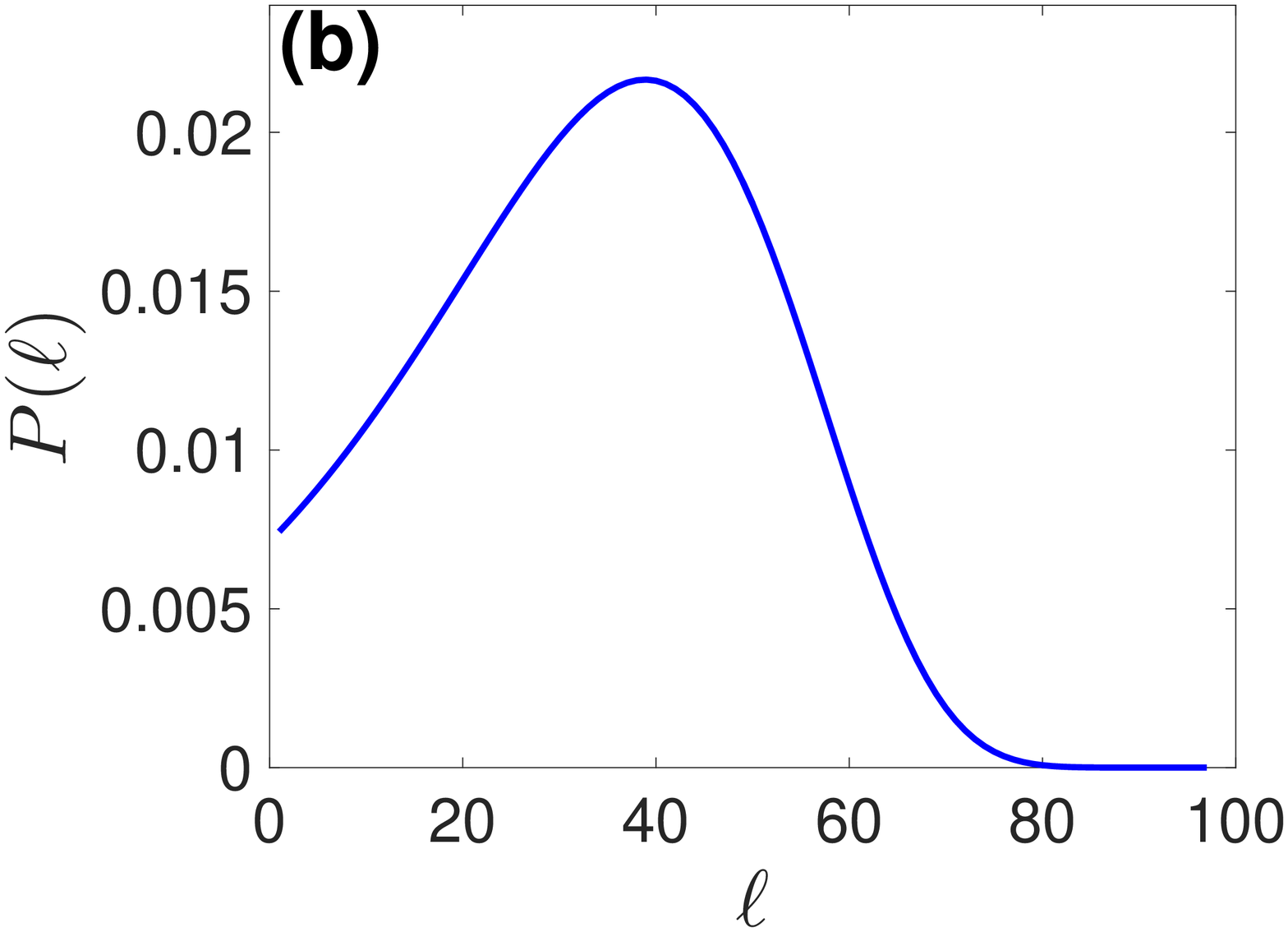}
}
\caption{
(a) The number of SAW paths, $n_T(\ell)$, 
vs. $\ell$ for 
an ER network of $N=100$ and $p=5/100$,
obtained from Eq.
(\ref{eq:nT(ell)}).
The peak is at $\ell \simeq 80$, in agreement with Eq.
(\ref{eq:lTpeak}).
(b) The distribution of SAW path lengths, $P(\ell)$, vs. $\ell$,
for the same network, obtained from Eq.
(\ref{eq:P(l)}).
The peak is at 
$\ell \simeq 40$,
in agreement with Eq.
(\ref{eq:mode1}). 
The difference between the locations of the two peaks reflects the fact
that long SAW paths are less likely to be pursued than shorter ones.
}
\label{fig:nP(ell)}
\end{figure}

In Fig. 
\ref{fig:nP(ell)}
we present the number of SAW paths, $n_T(\ell)$ of length $\ell$
(a) and the distribution of path lengths $P(\ell)$ (b) for an ER
network of size $N=100$ and $p=0.05$.
It is observed that the peak of $P(\ell)$ takes place at a smaller
$\ell$ value than the peak of 
$n_T(\ell)$.
This reflects the fact that the 
SAW paths which are actually pursued
are typically shorter than
SAW paths chosen at random from the list of all SAW paths.
The former SAW paths are often referred to as kinetic growth
self-avoiding walks
\cite{Herrero2007},
or true self-avoiding walks
\cite{Slade2011},
in contrast to the SAW paths which are uniformly
sampled among all possible self avoiding paths of a given lengths.

\begin{figure}
\centerline{
\includegraphics[width=7cm]{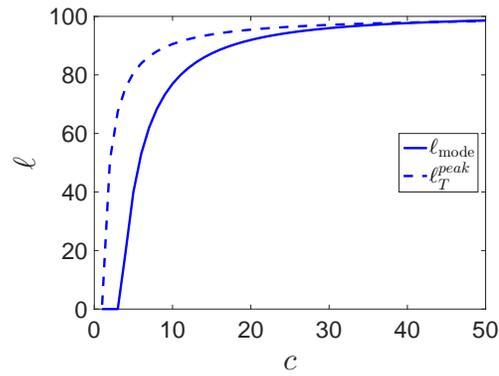}
}
\caption{
The location of the peaks, 
$\ell_T^{peak}$ 
(dashed line),
and
$\ell_{mode}$ 
(solid line)
of 
$n_T(\ell)$
and
$P(\ell)$,
respectively
vs. $c$.
Both curves increase monotonically as a function of $c$.
For small values of $c$,
$\ell_{mode}$ 
is much smaller than
$\ell_T^{peak}$,
which means that the huge number of long
SAW paths are rarely pursued.
}
\label{fig:ell(peak)}
\end{figure}

The reason that true self avoiding walks are typically shorter
is that the probability of an 
SAW path to be pursued is a decreasing function of 
its length, $\ell$.
More specifically, the number of paths
$n_T(\ell)$
proliferates at rate determined by $c(\ell)$,
and thus keeps increasing as long as $c(\ell)>1$.
On the other hand,
$P(\ell)$ 
is sensitive to the termination rate,
given by
$p_{\ell}(k=0)$,
which increases as a function of $\ell$.
Thus, $P(\ell)$ reaches its maximum earlier than
$n_T(\ell)$.
This is demonstrated in
Fig.
\ref{fig:ell(peak)},
in which
we present the locations of the peaks,
$\ell_T^{peak}$
and 
$\ell_{mode}$
as a function of $c$.
Both curves increase monotonically with $c$.
For dilute networks, it is shown that 
$\ell_{mode}$
is much smaller than
$\ell_T^{peak}$,
while for dense networks they are comparable.
A similar effect was observed long ago for SAWs on regular lattices
\cite{Amit1983}.

From another perspective, the distribution $P(\ell)$ follows the 
Gompertz law, where the termination rate increases with the number of
steps already pursued by the SAW.
Therefore, the fact that the number of available SAW paths, $n_T(\ell)$ 
increases with the length is overridden by the super-exponential decay
of the Gompertz distribution.

\section*{acknowledgements}

We thank Nathan Clisby for helpful comments.
%\newpage

\newpage

\end{document}